\documentclass[twocolumn, prl]{revtex4-2}
\usepackage[USenglish]{babel}
\usepackage{amsmath,amsfonts, amssymb, bm, colortbl, graphicx,hyperref,siunitx}
\usepackage[sort&compress,capitalize]{cleveref}
\newcommand{\cit}[1]{Ref.~\cite{#1}}

\begin{document}
\title{Transmission Line Circuit and Equation for an Electrolyte-Filled Pore of Finite Length}
\author{Mathijs Janssen}
\email{mathijsj@uio.no}
\affiliation{Department of Mathematics, Mechanics Division, University of Oslo, N-0851 Oslo, Norway}
\date{\today}

\begin{abstract}
I discuss the strong link between the transmission line (TL) equation and the TL circuit model for the charging of an electrolyte-filled pore of finite length.
In particular, I show how Robin and Neumann boundary conditions to the TL equation, proposed by others on physical grounds, also emerge in the TL circuit subject to a stepwise potential.
The pore relaxes with a timescale $\tau$, an expression for which consistently follows from the TL circuit, TL equation, and from the pore's known impedance.
An approximation to $\tau$ explains the numerically determined relaxation time of the stack-electrode model of Lian \textit{et al.} [\href{https://doi.org/10.1103/PhysRevLett.124.076001}{Phys. Rev. Lett. \textbf{124}, 076001 (2020)}].
\end{abstract}
\maketitle

\begin{figure}[b]
\includegraphics[width=0.85\linewidth]{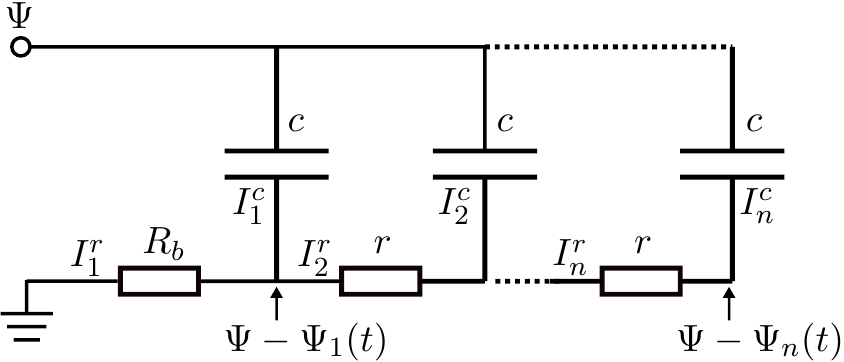}
\caption{TL circuit with $n$ capacitors of capacitance $c$, $n-1$ resistors of resistance $r$, and one resistor of resistance $R_{b}$. \label{fig1}}
\end{figure}

In the early 1960s, de Levie wrote two seminal papers on electric double layer formation in porous electrodes \cite{delevie1963, de1964porous}.
Both papers start with the transmission line (TL) circuit for an electrolyte-filled pore (\cref{fig1}), whose resistance $R$ and capacitance $C$ are distributed over many infinitesimally small resistors and capacitors.
From this circuit, de Levie argued that $\psi(z,t)$---the electrostatic potential difference between the pore's surface and center line at time $t$ and location $z$---follows the TL equation,
\begin{equation}\label{eq:TLequation}
RC \partial_{t}\psi=\ell^2 \partial_{z}^{2}\psi\,,
\end{equation}
where, for dimensional reasons, I introduced a length scale $\ell$, which is absent in Refs.~\cite{delevie1963, de1964porous}.
Both the TL circuit and TL equation found countless applications, particularly for the interpretation for electrochemical impedance spectroscopy experiments \cite{de1967electrochemical,bisquert2002,barsoukov2005impedance, newman2012electrochemical,conway2013electrochemical}.
With the ongoing interest in electrolyte-filled nanopores in general \cite{mirzadeh2014, tivony2018charging, martinez2019, gupta202charging, timur2020relation} and in nanoporous supercapacitors in particular \cite{biesheuvel2010, lian2020blessing, breitsprecher2020how}, de Levie's work is as relevant today as it was six decades ago.
Yet, while Refs.~\cite{delevie1963, de1964porous} considered \cref{eq:TLequation} on a semi-infinite interval $z=[0,\infty)$, more relevant for the dc response of supercapacitors is the TL equation on a finite interval, which was studied by Biesheuvel and Bazant \cite{biesheuvel2010} and more recently by Gupta, Zuk, and Stone \cite{gupta202charging}.
Here, I discuss the intimate relation between the TL circuit and the TL equation on a finite interval, by considering a finite-difference scheme of the latter.
In particular, the Robin and Neumann boundary conditions of Refs.~\cite{biesheuvel2010,gupta202charging}, proposed there on physical grounds, also emerge in the TL circuit itself.

The TL circuit in \cref{fig1} distributes $R$ and $C$ over $n-1$ resistors of resistance $r$ and $n$ capacitors of capacitance $c$, so that $R=r(n-1)$ and $C=cn$ \footnote{The circuit in \cref{fig1} does not account for Faradaic reactions on the electrode surface, typically modelled through a Faradaic impedance parallel to the capacitors \cite{conway2013electrochemical}.}.
A bulk electrolyte reservoir is represented in the circuit by a resistor of resistance $R_{b}$.
Now, the current from the $i$th capacitor reads $I^{c}_{i}(t)=c\dot{\Psi}_{i}(t)$ for $i=1,\ldots,n$, where $\dot{\Psi}_{i}(t)$ is the time derivative of the voltage $\Psi_{i}(t)$ across this capacitor. 
Kirchoff's junction rule gives $I^{c}_{i}(t)=I^{r}_{i}(t)-I^{r}_{i+1}(t)$ for $i=1,\ldots,n-1$ and $I^{c}_{n}(t)=I^{r}_{n}(t)$, with $I^{r}_{i}(t)$ the current through the $i$th resistor; 
Ohm's law states that $I^{r}_{i}(t) r = \Psi_{i-1}(t)-\Psi_{i}(t)$ for $i=2,\ldots,n$ and that $I^{r}_{1}(t) R_{b} = \Psi-\Psi_{1}(t)$, with $\Psi$ the potential of an external voltage source, suddenly applied at $t=0$. 
Writing $\mathbf{\Psi}(t)= \left[\Psi_{1}(t),\ldots, \Psi_{n}(t)\right]^{\intercal}$, $\mathbf{e}_{1}= \left[1,0,\ldots\right]^{\intercal}$, $\xi \equiv R/R_{b}$, and $\zeta \equiv r/R_{b}$ [hence, $\zeta = \xi/(n-1)$], I find 
\begin{subequations}\label{eq:TLcircuiteq}
\begin{align}
RC\dot{\mathbf{\Psi}}(t)&=n\xi \Psi \mathbf{e}_{1}+ n(n-1)M \mathbf{\Psi}(t)\,,\label{eq:matrixeq}\\
M&=
\begin{bmatrix}
-1-\zeta& 1& &&\\
1& -2& 1&& \\
&\ddots & \ddots &\ddots &\\
& & 1& -2&1\\
& & & 1&-1\\
\end{bmatrix}\,,\label{eq:matrix}
\end{align}
\end{subequations}
with $M\in \mathbb{R}^{n\times n}$ (cf.~\cit{lian2020blessing}). 
For initially uncharged capacitors [$\mathbf{\Psi}(0) = \mathbf{0}$], \cref{eq:TLcircuiteq} is solved by
\begin{equation}\label{eq:potentiostatic}
\frac{\mathbf{\Psi}(t)}{\Psi} = \zeta U \left[\exp{\!\left(\frac{Dn(n-1)t}{RC}\right)}-1\right] D^{-1}U^{-1} \mathbf{e}_{1}\,,
\end{equation}
where $D={\rm diag}(\lambda_{1},\ldots,\lambda_{n})$ contains the eigenvalues $\lambda_{i}$ of $M = UDU^{-1}$, which are all negative. 

Consider now a cylindrical pore of length $\ell$ and radius $a$ with the same resistance $R$ and capacitance $C$ as the TL circuit above, subject to the same instantaneous potential $\Psi$.
The pore is closed at $z=\ell$ and in contact with a bulk reservoir of resistance $R_{b}$ at $z=0$.
I study $\psi(z,t)$ in this pore through the TL equation~\eqref{eq:TLequation} subject to Robin and Neumann boundary conditions,
\begin{subequations}\label{eq:TLinitandbcs}
\begin{align}
\psi(z,0)&=0\,, &\quad \quad &z\in[0,\ell]\,,\label{eq:TLinit}\\
\ell\partial_{z}\psi(0,t)&=\xi[\psi(0,t)-\Psi], &\quad \quad &\partial_{z}\psi(\ell,t)=0\,.\label{eq:TLbcs}
\end{align}
\end{subequations}
Reference~\cite{biesheuvel2010} proposed a similar Robin condition at $z=0$ on the basis of $\psi(z)$ being linear in the reservoir ($z<0$);
\cit{gupta202charging} refined the same argument for a pore with overlapping electric double layers, that is, when the Debye length is comparable to the pore's radius $\lambda_{D}\approx a$.
For that case, $\psi(z,t)$ should not reach $\Psi$ at late times, and the TL circuit must be adopted accordingly \cite{gupta202charging}. 

The solution to \cref{eq:TLequation,eq:TLinitandbcs} reads \cite{beck1992heat}
\begin{subequations}\label{eq:beckcole}
\begin{align}
\frac{\psi(z,t)}{\Psi}&=1-\sum_{j\ge1}\frac{4\sin \beta_j \cos \left[\beta_j\left(1-z/\ell\right)\right]}{2\beta_j +\sin 2\beta_j} 
\exp{\!\left[-\frac{\beta_j^2 t}{RC}\right]}\,,\label{eq:beckcole2}
\intertext{where $\beta_j$ with $j=1,2,\ldots$ are the solutions of the transcendental equation }
&\beta_j \tan \beta_j = \xi\,.\label{eq:continuumtranscendental}
\end{align}
\end{subequations}

For comparison, I also mention the solution to \cref{eq:TLequation} on a semi-infinite slab $z\in[0,\infty)$ subject to the same Robin condition at $z=0$ \cite{carslaw1959conduction},
\begin{align}\label{eq:deLevie}
\frac{\psi(z,t)}{\Psi} &=-\exp\left[\xi \frac{z}{\ell}+\xi^2 \frac{t}{RC}\right] \mathrm{erfc}\left[\sqrt{\frac{z^2}{\ell^2}\frac{ RC}{4t}}+\xi\sqrt{\frac{t}{RC}}\right]\nonumber\\
&\quad+\mathrm{erfc}\sqrt{\frac{z^2}{\ell^2}\frac{ RC}{4t}}\,.
\end{align}

Note that $R_b$ entered the TL equation through the Robin condition \cref{eq:TLbcs}.
For $\xi=R/R_b\to\infty$, this Robin condition simplifies to de Levie's Dirichlet condition \cite{delevie1963}.
In this limit, \cref{eq:continuumtranscendental} simplifies to $\cos \beta_{j}=0$---solved by $\beta_{j}=(j-1/2)\pi$---and $\psi(z,t)$ simplifies accordingly.
Meanwhile, only the last term of \cref{eq:deLevie} remains for $\xi\to\infty$ and $\psi(z,t)$ reduces to Eq.~9 of \cit{delevie1963}.

\begin{figure} 
\includegraphics[width=\linewidth]{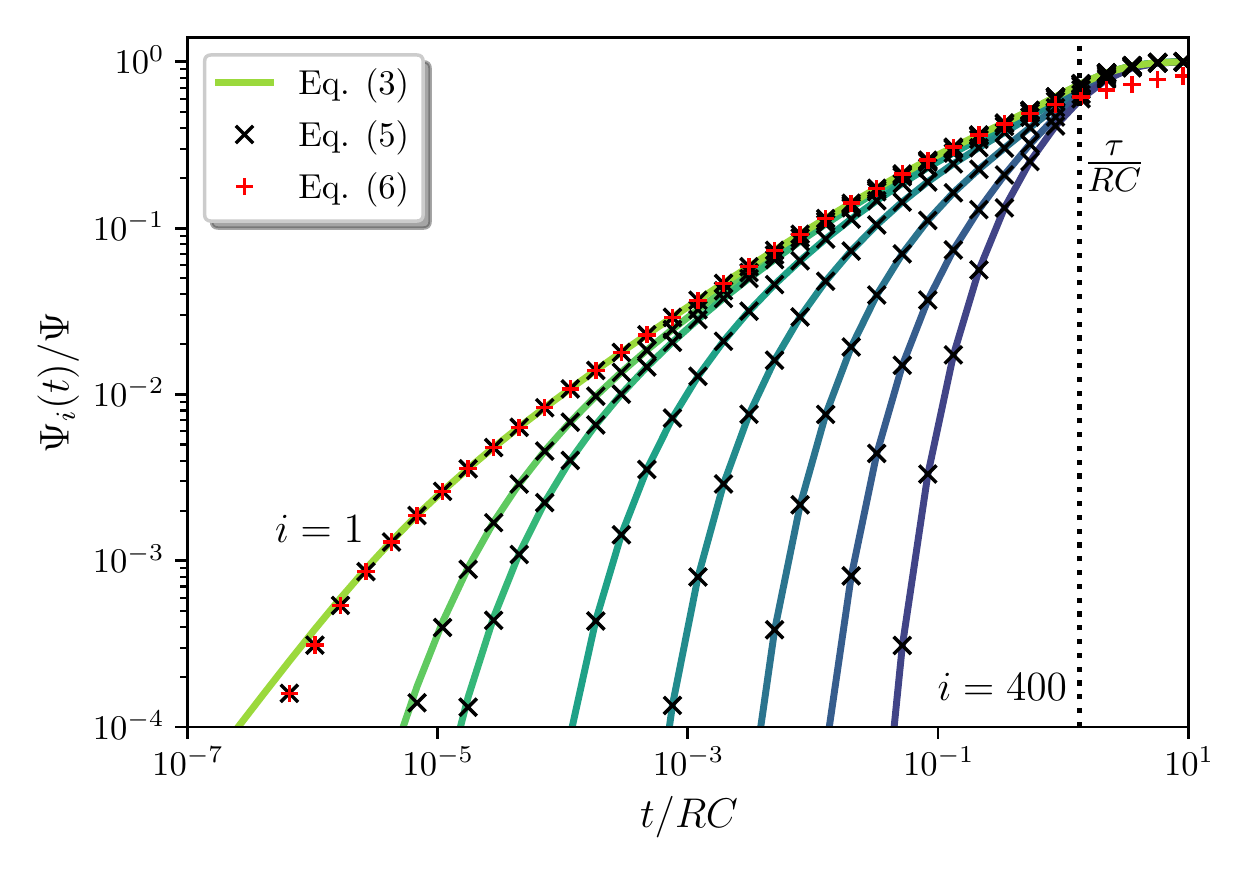}
\caption{TL-circuit potential drops $\Psi_{i}(t)$ [\cref{eq:potentiostatic}, lines] and TL-equation solutions $\psi\bm{(}z/\ell=(i-1/2)/n,t\bm{)}$ [\cref{eq:beckcole}, crosses] and $\psi\bm{(}z=\ell/(2n),t\bm{)}$ [\cref{eq:deLevie}, pluses], all divided by $\Psi$, for $i=(1,3,5, 15, 40,100,200,400)$, $n=400$, and $\xi=R/R_{b}=1$. 
The sum in \cref{eq:beckcole2} is truncated after ${\rm max}(j)=1000$. 
The dotted line indicates the late-time relaxation time $\tau=RC/\beta_{1}^2$. \label{fig2}}
\end{figure}
\Cref{fig2} shows $\Psi_{i}$ [\cref{eq:potentiostatic}, lines] and $\psi\bm{(}z=(i-1/2)\ell/n,t\bm{)}$ [\cref{eq:beckcole}, crosses] for $R_{b}=R$, $n=400$, and $i=(1,3,5, 15, 40, 100, 200, 400)$. 
As $\Psi_{i}$ describes the potential drop between the pore's surface and centerline along the pore from $z=(i-1)\ell/n$ to $z=i\ell/n$, I evaluate $\psi$ at the center of this interval. 
\Cref{fig2} shows that predictions from \cref{eq:potentiostatic,eq:beckcole} agree well, except for $i=1$ and $t/RC<10^{-6}$.
For comparison, \cref{fig2} also shows $\psi\bm{(}z=\ell/(2n),t\bm{)}$ from \cref{eq:deLevie} (pluses).
Predictions from \cref{eq:beckcole,eq:deLevie} coincide up to $t\approx RC$, when the potential perturbations reach $z=\ell$ and, hence, the Neumann condition in \cref{eq:TLbcs} becomes important.

To better understand {\it why} \cref{eq:potentiostatic,eq:beckcole} agree so well, I turn to a finite-difference description of \cref{eq:TLequation,eq:TLinitandbcs}.
Following \cit{strang2014functions}, I discretise $z$, but not $t$.
Partitioning $[0,\ell]$ into $m-1$ intervals of width $h = \ell/(m-1)$ yields a uniform grid of $m$ grid points, at $z_{k}=kh$ with $k\in\{0,\ldots,m-1\}$.
On these grid points, the continuous electrostatic potential is approximately $\psi_{k}=\psi(z_{k})$.
A central difference approximation now gives $\partial_{z}^{2}\psi(z_{k})\simeq (\psi_{k-1}-2\psi_{k}+\psi_{k+1})/h^2$.
To implement the Robin boundary condition at $z=0$, I introduce a ghost grid point at $z=-h$ and corresponding $\psi_{-1}$.
Now, approximating the $z$ derivative through a backward difference $\partial_{z}\psi(0)\simeq (\psi_{0}-\psi_{-1})/h$, the Robin boundary condition yields $\psi_{-1}=\psi_{0}+\xi(\Psi-\psi_{0})/(m-1)$. 
Similar reasoning and a forward difference yields for the Neumann condition that $\psi_{m}=\psi_{m-1}$ \cite{strang2014functions}.
After grouping the above expressions and writing $\boldsymbol{\psi}(t)= \left[\psi_{1}(t),\ldots, \psi_{m-1}(t)\right]^{\intercal}$, \cref{eq:TLequation,eq:TLinitandbcs} are approximated by
\begin{equation}\label{eq:findiff}
RC \dot{\boldsymbol{\psi}}(t)=(m-1)\xi \Psi \mathbf{e}_{1}+ (m-1)^2 M \boldsymbol{\psi}(t)\,,
\end{equation}
with $M\in \mathbb{R}^{m\times m}$ as in \cref{eq:matrix}.
Setting $m=n$, \cref{eq:findiff,eq:TLcircuiteq} are very similar: the prefactors on their right-hand sides contain differences that are of subleading order in $n$. 
Indeed, replotting \cref{fig2} for smaller $n$, I observed that differences between \cref{eq:potentiostatic,eq:beckcole} became larger, while for $n>500$, both methods were practically indistinguishable (not shown).
Note, first, that the differences between \cref{eq:potentiostatic,eq:beckcole} are unrelated to the truncation of \cref{eq:beckcole} at finite $j$:
my numerical observation that this sum was converged is reinforced by the overlap of \cref{eq:beckcole,eq:deLevie} at early times.
Second, note that differences between \cref{eq:findiff,eq:TLcircuiteq} of subleading order in $n$ could not be circumvented altogether, for instance, by changing the TL circuit or the above finite-difference scheme: the order of $M$ in \cref{eq:matrixeq} is equal to the number of capacitors in the circuit, which also sets the factor $n$ in $n(n-1)M$.
Conversely, in \cref{eq:findiff}, the order of $M$ is given by the number of grid points, while the prefactor of $M$ is set by the number of intervals, which is always one smaller.
Lastly, differences between \cref{eq:findiff,eq:TLcircuiteq} being of subleading order in $n$ means that those equations are equal in the limit $n\to\infty$.
Thus, different from the physical arguments of Refs.~\cite{gupta202charging,biesheuvel2010}, both the Robin and the Neumann boundary condition in \cref{eq:TLbcs} also emerge naturally in the TL circuit and \cref{eq:TLcircuiteq}, which governs its relaxation.

Important for applications of porous electrodes, \cref{fig2} suggests that $\psi(z,t)$ relaxes with a single late-time relaxation time, denoted $\tau$, throughout the pore.
This observation stands in contrast to de Levie's solution to the TL equation on a semi-infinite interval---the last term of \cref{eq:deLevie}---which relaxes with a position-dependent relaxation time $(z/\ell)^2 RC/4$ \cite{delevie1963}.
From \cref{eq:beckcole} it follows that $\tau=RC/\beta_{1}^2$, with $\beta_{1}$ the smallest solution to \cref{eq:continuumtranscendental}.
For example, $\xi=1$ yields $\tau/RC\approx 1.35 $, shown with a dotted line in \cref{fig2}.
Conversely, by the above-mentioned simplification of \cref{eq:continuumtranscendental}, $\xi\to\infty$ yields $\tau/RC=4/\pi^2$ \cite{mirzadeh2014,timur2020relation}.

The same relaxation behavior follows from the TL circuit:
as all eigenvalues $\lambda_{i}$ of $M$ are negative, it follows from \cref{eq:potentiostatic} that $\mathbf{\Psi}(t) $ relaxes at late times with the timescale
\begin{equation}\label{eq:tauplus}
 \tau = -\frac{RC}{n(n-1)\lambda_{+}}\,,
\end{equation}
with $\lambda_{+}=\max\{\lambda_{1},\ldots, \lambda_{n}\}$ the least negative eigenvalue of $M$.
For matrices of $M$'s form, the different $\lambda_{i}$ satisfy 
\begin{align}\label{eq:Q}
&U_{n}\left(\frac{\lambda_{i}}{2}+1\right)-U_{n-1}\left(\frac{\lambda_{i}}{2}+1\right)=\nonumber\\
&\quad \quad \left(1-\zeta\right)\left[U_{n-1}\left(\frac{\lambda_{i}}{2}+1\right)-U_{n-2}\left(\frac{\lambda_{i}}{2}+1\right)\right]\,,
\end{align}
where $U_{n}$ are $n$th degree Chebyshev polynomials of the second kind \footnote{$T_n(a=1,b=1-1/\zeta)$ in Eq. (1.1) of \cit{fonseca2007} equals $-M$ [\cref{eq:matrix}]. Notably, Eq.~(2.1) of \cit{fonseca2007} yields not the spectrum of $-T_{n}^{-1}/a$, as suggested there, but of $-T_{n}/a$.}.
With $U_{n}\left(\cos \vartheta\right)=\sin\bm{(}\left(n+1\right)\vartheta\bm{)}/\sin \vartheta $, inserting $\lambda_{i}=2\left[\cos (\vartheta_{i})-1\right]$ into \cref{eq:Q} yields, 
\begin{equation}\label{eq:sinexpression}
\frac{\sin\bm{(}\left(n+1\right)\vartheta_{i}\bm{)}-\sin\left(n \vartheta_{i}\right)}{1-\zeta}=
\sin\left(n \vartheta_{i}\right)-\sin\bm{(}\left(n-1\right)\vartheta_{i}\bm{)}\,.
\end{equation}
Using $\sin(\alpha\pm\beta)= \sin(\alpha)\cos(\beta)\pm\sin(\beta)\cos(\alpha)$ and dividing both sides of \cref{eq:sinexpression} by $\sin(n\vartheta_{i})\sin \vartheta_{i}$ yields
\begin{align}\label{eq:transcendental}
\frac{2-\zeta}{\zeta}\tan\left(n\vartheta_{i}\right) = \frac{\sin \vartheta_{i}}{1-\cos \vartheta_{i}}\,.
\end{align}
The smallest solution $\vartheta_{-}$ to \cref{eq:transcendental}, required to find $\lambda_{+}=2\left[\cos (\vartheta_{-})-1\right]$, lies in the interval $\vartheta_{-}\in[0,\pi/(2n)]$.
Thus, for $n\gg1$, one has $\vartheta_{-}\ll1$ and thus $\lambda_{+} =- \vartheta_{-}^2+O(\vartheta_{-}^4)$.
Now, for $n\gg1$ and provided that $\xi/n\ll1$, \cref{eq:transcendental} reduces to
\begin{equation}\label{eq:transcendental2}
n\vartheta_{-}\tan\left(n\vartheta_{-}\right)=\xi+O\left(n^{-1}\right)\,.
\end{equation}
Hence, for $n\gg1$, the late-time relaxation times of the TL circuit and the TL equation are governed by the same transcendental equation [\cref{eq:continuumtranscendental,eq:transcendental2}]. 

A Pad\'{e} approximation of order $[1/2]$ of the $\tan(n\vartheta_{-})$ term in \cref{eq:transcendental2} yields the approximate solution \cite{Wu_2018}
\begin{equation}
n\vartheta_{-}\approx\sqrt{\frac{3\xi}{3+\xi}}\,.
\end{equation}
From \cref{eq:tauplus} now follows the late-time response of the TL circuit as 
\begin{equation}\label{eq:tau_approx}
 \tau\approx \frac{1}{3}RC+R_{b}C\,. 
\end{equation}
This expression is inaccurate for small $R_b$: in the limit $\xi\to\infty$, the TL circuit expression \cref{eq:transcendental2} simplifies to $\cos\left(n\vartheta_{-}\right)=0$.
As anticipated, its solution $\vartheta_{-}=\pi/(2n)$ yields $\tau=4RC/\pi^2$, suggesting that the factor $1/3$ in \cref{eq:tau_approx} should be replaced by $4/\pi^2\approx0.41$,
\begin{equation}\label{eq:tau_approx2}
\tau\approx \frac{4}{\pi^2}RC+R_{b}C\,.
\end{equation}

\begin{figure}
\includegraphics[width=\linewidth]{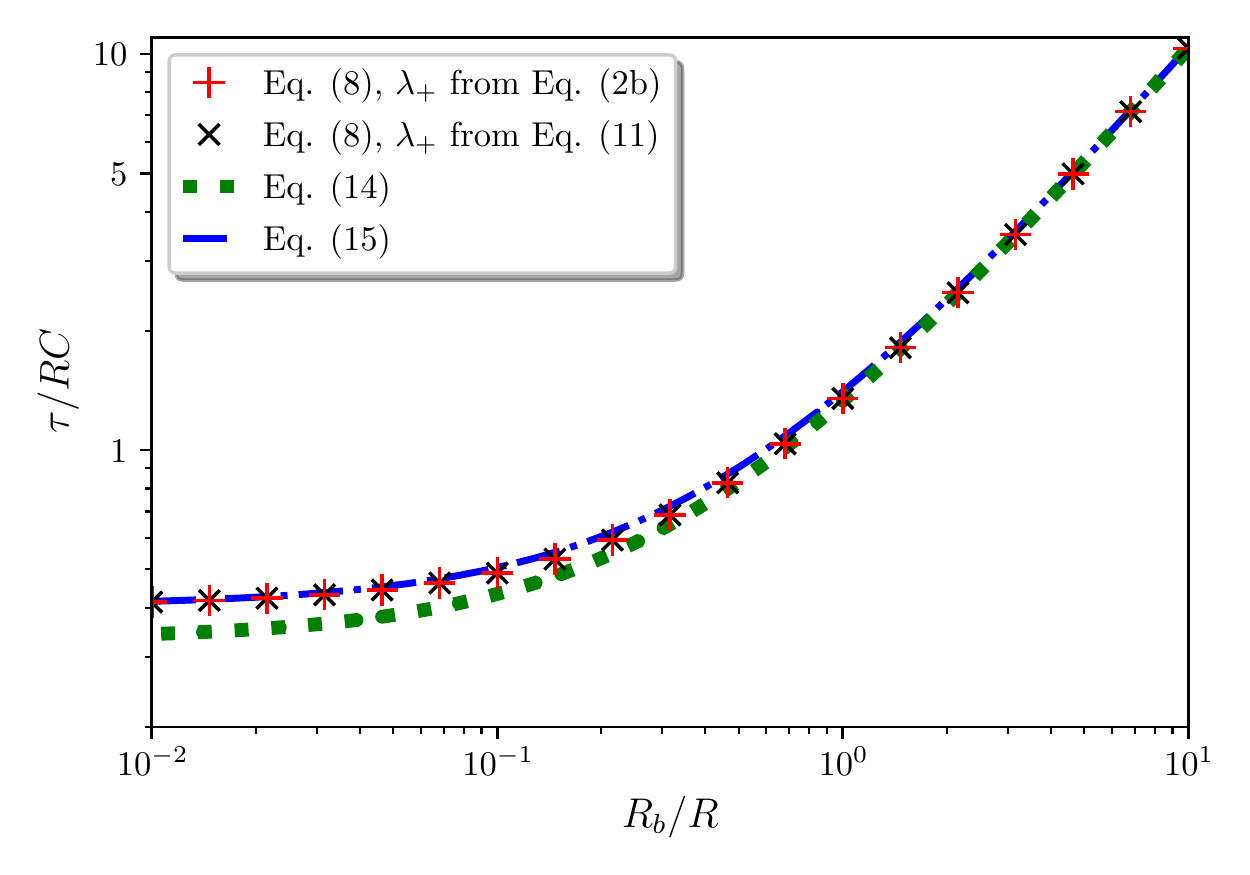}
\caption{Late-time relaxation time $\tau$ [\cref{eq:tauplus}] for $n=100$, with $\lambda_{+}$ determined numerically from $M$ (pluses) and \cref{eq:transcendental} (crosses). 
Also shown are the approximate solutions \cref{eq:tau_approx} (dotted line) and \cref{eq:tau_approx2} (dash-dotted line). \label{fig3}}
\end{figure}
\Cref{fig3} shows $\tau$ [\cref{eq:tauplus}] with $\lambda_{+}$ determined from $M$ directly (red pluses) and from \cref{eq:transcendental} (black crosses) as well as the approximations \cref{eq:tau_approx} (dotted line) and \cref{eq:tau_approx2} (dash-dotted line).
Since crosses and pluses overlap, \cref{eq:transcendental} successfully captures $\lambda_{+}$.
As expected, \cref{eq:tau_approx} accurately approximates $\tau$ for $R_{b}/R\gg1$ but not for $R_b/R  \lessapprox  1$.
Conversely, \cref{eq:tau_approx2} is in excellent agreement with \cref{eq:tauplus} at both $R_{b}/R\ll1$ and $R_{b}/R\gg1$ but slightly less so around $R_{b}/R\approx1$.

There is yet another route to the timescale $\tau$ with which a finite-length pore relaxes in response to a stepwise potential: through its known impedance $Z(\mathrm{i}\hspace{0.2mm}\omega)=\sqrt{R/(\mathrm{i}\hspace{0.2mm} \omega C)}\coth\sqrt{\mathrm{i}\hspace{0.2mm} \omega RC}$ \cite{de1967electrochemical}. 
Here, $\mathrm{i}=\sqrt{-1}$ and $\omega$ is the angular frequency of a sinusoidal potential applied to the pore. 
At low frequencies $\hat{Z}(s\approx0)\approx \hat{Z}_{l}(s)=R/3+1/(Cs)$, where the complex frequency $s$ appears instead of $\mathrm{i}\hspace{0.2mm}\omega$ \cite{bisquert2002, barsoukov2005impedance}. 
The same $\hat{Z}_{l}(s)$ applies to a series connection of a resistor of resistance $R/3$ and a capacitor of capacitance $C$. 
To account for the bulk with which the pore is in contact, I add a resistor of resistance $R_{b}$ in series with these two elements.
Subjecting this circuit to a step potential $V(t)=V_{0}\Theta(t)$, with $\Theta(t)$ the Heaviside function, drives a current $I(t)=\mathcal{L}^{-1}\{\hat{V}(s)/[\hat{Z}_{l}(s)+R_b]\}\propto \exp[-t/\tau]$, with $\mathcal{L}^{-1}$ the inverse Laplace transform, $\hat{V}(s)=\mathcal{L}\{V(t)\}=V_{0}/s$, and $\tau$ precisely as in \cref{eq:tau_approx}. 
Yet, inverse Laplace transformations of approximate expressions yield wrong relaxation times if the original function has different poles than its approximation \cite{janssen2018}. 
Such is the case for $1/\hat{Z}_{l}(s)$.
The exact current $I(t)=\mathcal{L}^{-1}\{\hat{V}(s)/[\hat{Z}(s)+R_b]\}$ relaxes at late times with $\tau=-1/s^{*}$, with $s^{*}$ the first solution to $\sqrt{R/(sC)}\coth\sqrt{s RC}+R_b=0$ on the negative $s$ axis. 
Substituting $sRC=-\beta_j^{2}$, we recover \cref{eq:continuumtranscendental}; hence, $I(t)$ relaxes precisely as $\psi(z,t)$ in \cref{eq:beckcole2}.

While several papers included a bulk resistance in the TL circuit \cite{de1964porous, biesheuvel2010, gupta202charging}, the influence of $R_{b}$ on the relaxation of the TL circuit is not generally recognized, \cit{kroupa2016modelling} being a notable exception.
The often-used TL timescale $\lambda_{D}\ell^2/(Da)$ \cite{biesheuvel2010, mirzadeh2014, tivony2018charging, gupta202charging}, with $D$ the ionic diffusivity, does not account for $R_b C$, nor for $RC$'s  prefactors in \cref{eq:tau_approx,eq:tau_approx2}.
Hence, depending on the geometry of interest, particularly on the distance of the pore to a counter electrode, a pore's relaxation time can deviate significantly from $\lambda_{D}\ell^2/(Da)$.
Still, in electrodes with ultranarrow pores---much beyond the validity of the TL equation---attenuation of the in-pore diffusivity probably yields $R\gg R_{b}$, making pore entrance the rate-limiting step of electrode charging \cite{breitsprecher2020how}.

As a corollary, I show how \cref{eq:tau_approx} sheds light on the recently proposed stack-electrode model for supercapacitor charging \cite{lian2020blessing}.
In this model, a porous electrode of thickness $H$ was represented by a stack of $n$ flat, metallic yet permeable sheets of area $A$, with a constant spacing $h$, so that $r = h \lambda_{D}^2/(\varepsilon D A)$ and $c = 2\varepsilon A/\lambda_D $. 
Two such electrodes were in contact with a bulk of length $2L$; hence, $R_{b}/R=L/H$.
\Cref{eq:tau_approx} now yields $\tau = (2+2H/3L)n\lambda_{D}L/D$, which, for large $n$, is in reasonable agreement with the fitted timescale 
$\tau_{n}= [\left(2+0.75H/L\right)n-1-0.91H/L]\lambda_{D}L/D$ of \cit{lian2020blessing}.
While both $\tau_{n}$ and $\tau$ from \cref{eq:tau_approx} are based on approximations, differences between them must also stem from the different $n$th sheet in the stack-electrode model, which had half the capacitance of the other sheets.
As $\tau_{n}$ captured the short timescale of the biexponential current decay in the experiments of \cit{janssen2017coulometry}, $\tau$ as calculated here accurately describes the same timescale as well 
\footnote{A stagnant diffusion layer is sometimes introduced in lieu of the bulk reservoir length $2L$ \cite{biesheuvel2010, gupta202charging}. 
This choice would yield $\tau$ in poor agreement with \cit{janssen2017coulometry}.}.
The stack-electrode model also captured the second, larger timescale of the transient current measured in \cit{janssen2017coulometry} and ascribed it to the \SI{0.1}{\volt} applied there---large, compared to the thermal voltage of \SI{24}{\milli\volt}.
Such potentials fall outside the region of validity of the TL equation \cite{biesheuvel2010}.

Concluding, I have exposed the intimate relation between the TL circuit model for a pore in contact with an electrolyte reservoir and the TL equation subject to Robin and Neumann boundary conditions.
The pore relaxes with a $R_{b}/R$-dependent relaxation time that explains one of the two dominant relaxation timescales of Refs.~\cite{lian2020blessing,janssen2017coulometry}.

I thank Carlos da Fonseca and Cheng Lian for inspiring discussions, Christian Pedersen and Stephane Poulain for useful comments on this work, and an anonymous referee for pointing out $\hat{Z}_{l}(s)=R/3+1/(Cs)$ to me.
The research leading to these results has received funding from the European Union's Horizon 2020 research and innovation programme under the Marie Sk\l{}odowska-Curie grant agreement No 801133.

\end{document}